# Two Models of Mind Blanking

Angelica Kaufmann[1,4], Sara Parmigiani[2,4], Toshikazu Kawagoe[3], Elliot Zabaroff[4], Barnaby Wells[4]

**Abstract**

Mind blanking is a mental state in which attention does not bring any perceptual input into conscious awareness. As this state is still largely unexplored, we suggest that a comprehensive understanding of mind blanking can be achieved through a multifaceted approach combining self-assessment methods, neuroimaging, and neuromodulation. In this article, we explain how EEG and TMS could be combined to help determine whether mind blanking is associated with a lack of mental content or a lack of linguistically or conceptually determinable mental content. We also question whether mind blanking occurs spontaneously or intentionally and whether these two forms are instantiated by the same or different neural correlates.

**keywords**
mind blanking; attention; perception; meta-awareness; self-report; Broca's area; conscious awareness; mind-wandering.

## 1. What is mind blanking?

Conscious experiences are typically associated with meta-awareness when, at a given time, we exercise the ability to take explicit note of the current contents of our thoughts (Schooler et al., 2011; Dunne et al., 2019). However, there are moments when our thoughts seem to be empty, or their content is not accessible. It remains debatable what qualifies as exercising meta-awareness about these seemingly contentless or inaccessible mental states, whether these exist in the first place, or whether these amount to mental states the content of which is merely not linguistically or conceptually accessible (as hypothesized by Kawagoe et al., 2019, Fell, 2022). In addition, exercising meta-awareness about such mental states may also be determined by whether these states are intentional or spontaneous occurrences (Boulakis et al., 2023bioRxiv). In this opinion article, we explore potential measurement techniques that could be employed to further study these mental states during which attention calls no perceptual input into conscious awareness. We refer to this phenomenon as mind blanking.

[1] Ruhr-Universität Bochum, Institut für Philosophie II, Universitätsstr. 150, D-44780 Bochum, Germany.
[2] Department of Psychiatry and Behavioral Sciences, Stanford University Medical Center, Stanford, CA, 94305, USA.
[3] School of Humanities and Science, Kyushu Campus, Tokai University, Higashi-Ku, Toroku 9-1-1, Kumamoto, 862-8652, Japan.
[4] Mind and Cognition Lab, PhiLab, University of Milan, via Festa del Perdono 7, 20122 Milan, Italy.

Research on mind blanking often finds a home within the study of mind-wandering (Seli et al., 2016). However, it is important to treat mind blanking as a distinct mental state. For example, it is an experience familiar to many that of finding ourselves caught absorbed in thoughts about something different from, and unrelated to, what we are doing at a given time of the day, such as driving the car and simultaneously thinking about the movie we watched last night or the plans we have for next weekend. These are instances of mind-wandering, a phenomenon consisting of times of the day during which what we mentally attend to is unrelated to what we are doing, perceiving, or experiencing emotionally (see Mittner, 2016; Irving and Glasser, 2019), barring all cases in which we are deliberately engaging in forms of mental action, such as mentally solving a math problem or planning what to cook for a dinner party. During mind wandering our attention is decoupled from perception and conscious awareness. We engage in a lot of these quasi-perceptual states in the form of mental imagery (Nanay 2015, 2018, 2021). Recent literature has emphasised the multifaceted nature of mind-wandering, including its conscious and intentional aspects (see Robison et al., 2020; Unsworth et al., 2018; Unsworth et al., 2020). We believe comparable and distinct room for research should be dedicated to its neighbouring phenomenon, mind blanking (Ward and Wegner, 2013). Unlike mind-wandering, during mind blanking experiences, our mental life appears either experientially empty or, this is still under investigation, inaccessible at a meta-conscious or linguistic level., This hypothesis is supported by findings that pinpoint the deactivation of Broca's area and parts of the default mode network (specifically, the hippocampus) as the neural correlates of mind blanking (Kawagoe et al., 2019). One further way to explain why mind blanking is different from mind wandering can be found in the distinction that many neuroscientists and philosophers hold between access consciousness and phenomenal consciousness (introduced by Block, 1995).

Access consciousness is understood as the cognitive ability to process information available for verbal reports, reasoning, and the guidance of behaviour. This deals primarily with the accessibility of representations and how they can be used to guide decision-making and action. Whereas, phenomenal consciousness refers to the subjective experience of being aware of something. It is the qualitative aspect of consciousness that allows us to have sensory experiences such as seeing colours, hearing sounds, and feeling emotions. While access consciousness is concerned with the functional aspects of consciousness, such as how information is processed and used, phenomenal consciousness is more concerned with the subjective experience of consciousness itself. This distinction is relevant for the present context because these dimensions of consciousness can be dissociable, as information may inhabit one's phenomenal consciousness without necessarily being accessible for cognitive appraisal or verbalisation.

While mind-wandering grants us access to mental content and makes us conscious of what we mind-wander about (e.g., we *have* access to the content of our thoughts about our plans for the weekend that occur spontaneously while we are driving the car), mind-blanking does not. During mind blanking, the content of our thoughts is either verbally inaccessible or non-existing and we have limited or no *access* to our conscious experience. We are however *phenomenally* conscious of this inaccessibility or lack of content. Mind blanking is thus a subjective experience in the phenomenal sense, but we are unable to report on it.

This parallelism between mind wandering and mind blanking in relation to the access versus phenomenal consciousness distinction is particularly relevant to the model of mind blanking according to which this experience involves a lack of verbal content. According to this model, a car driver may be phenomenally conscious of driving and this could be a mind blanking experience if there is no accompanying stream of verbal thought. The awareness that this experience was one of mind blanking occurs a posteriori as a result of metacognitive processes. However, the parallelism may not be suitable for the model according to which mind blanking is a contentless state because we lack evidence for assessing how this model could be associated with any phenomenological experience. This is one additional reason why further research into the two models is needed. It is important to notice that, long before neuroscience could enter into this debate, James (1980) wrote that conscious experiences could be experienced as continuous despite containing "interruptions, time-gaps during which consciousness goes out altogether to come into existence again at a later moment...". What James referred to as 'interruptions', aligns with the contemporary concept of mind blanking. Whilst James described mind-blanking as a consciousness black-out, it is now recognised that not all conscious experiences are accessible, yet we experience them.

The view of a constantly thought-oriented mind during wakefulness has been challenged by research on mind blanking (see, Ward and Wegner, 2013), and more recently by studies on the functions of slow waves, which are highly relevant to the methodology we are suggesting (Andrillon et al., 2019; Andrillon et al., 2021; Mortaheb et al., 2022). Even though our focus is not on slow waves, the work of Andrillon et al. 2021 suggests that mind blanking manifests as attentional lapses and that it can be explained by the occurrence of sleep-like, low-frequency, high-amplitude waves. The regional variations in slow wave occurrences seem to predict the nature and behavioural attributes of these attentional lapses. This has led us to incorporate EEG data collection in our study methodology to investigate whether we can observe a scalp topography resembling that reported by Andrillon et al. 2021 and Mortaheb et al., 2022.

Moreover, research by Kawagoe, Onoda and Yamaguchi, 2019 indicates that mind blanking could be attributed to an inability to report mental content. In this paper, we explore the implications of these findings and propose potential avenues for further investigation.

We argue that a comprehensive understanding of mind blanking can be achieved through a multifaceted approach, integrating self-assessment techniques with neuroimaging and neuromodulation methods.

Accordingly, we suggest how to combine EEG and TMS to investigate an aspect of mind blanking. Among the various questions surrounding mind blanking, we selectively address one with the aim not to provide definitive answers but to stimulate the debate and the research methodologies that could be employed to address it: we seek to discern whether mind blanking amounts to a lack of mental content or to a lack of linguistically or conceptually determinable mental content, which may be due to the silencing of inner speech caused by the deactivation of Broca's area.

Additionally, we evaluate this question in the context of instances when mind blanking occurs spontaneously versus when it is induced intentionally, and investigate whether these two forms of mind blanking are underpinned by the same or different neural correlates.

## 2. Two models discussed

Mind blanking is understood by some researchers as a sudden interruption in the stream of consciousness (Gennaro, Hermann, and Sarapata, 2006), while others describe it as an awareness of experiencing a mental void, characterized by an absence of any content in consciousness along with a sense that the retrieval of any information is not imminent (see the work of Efklides, 2014; Efklides and Touroutoglou, 2010; Moraitou and Efklides, 2009). Both perspectives regard mind blanking as a particular type of memory failure. Similarly, Bargh and Chartrand (1999) characterise mind blanking as a lack of conscious awareness, stating that during these episodes of 'blankness', as they call them, individuals are not aware of any stimuli, whether internal or external. Interestingly, they note that this portrayal of the phenomena might imply that mind blanking would cause one to lose control of their actions. However, if this were the case, the same should apply to mind-wandering, as we said before while illustrating the car driving example. But the car driving example tells us that mind-wandering does not necessarily result in a loss of control over one's actions (Pepin et al., 2020). As more recent works suggested, conscious awareness is unnecessary for much of human functioning; rather, most cognitive processing and behavioural control seems to occur outside of conscious awareness (Metzinger, 2013). For example, Metzinger (2020) introduced minimal phenomenal experience. This notion refers to the most basic form of conscious awareness, characterized by a simple, unadorned sensory experience that lacks complex cognitive and reflective elements. It is an attenuated form of consciousness where higher-order thinking, self-awareness, and elaborate conceptualizations are absent, but there still remains a fundamental level of sensory awareness. In contrast, mind blanking is best described as a state where there is a complete absence of conscious content. This implies that during mind blanking, not only there is an absence of complex thoughts, but also an apparent absence of any form of awareness, including the basic sensory awareness that characterizes minimal phenomenal experience.

To elucidate the distinction, let's revisit the example of driving a car. During minimal phenomenal experience, we might still be aware of the tactile sensations of the steering wheel, the visual input of the road, and the auditory stimuli from the car's engine, but without reflective or elaborate thoughts accompanying these sensations. Consciousness is pared down to its essential sensory components. However, in a state of mind blanking, even this basic sensory awareness might appear to be absent. A driver might arrive at their destination with no recollection or awareness of any part of the journey, including the sensory experiences. It is as if there was a temporary interruption in conscious experience altogether.

An additional illustrative example can be observed in the experience of floating in a sensory isolation float tank. In a float tank, external stimuli are minimized to such an extent that individuals often report losing the sense of where their body ends and where the water begins. This sensory deprivation can sometimes lead to an experience that is akin to mind blanking, where individuals feel they lack thoughts or mental content. According to Metzinger's view, this can be interpreted as consciousness retracting to its most basic form. The individual in the float tank is still conscious but in a highly reduced state, devoid of sensory input, cognitive processing, or self-referential thoughts. This distilled state of consciousness aligns with the notion of minimal phenomenal experience, suggesting that during the seeming emptiness of mind blanking, a fundamental, yet sparse, conscious presence persists. Mind blanking, on the other hand, would imply an absence of awareness even of these basic sensory experiences. While minimal phenomenal experience represents consciousness in its most rudimentary form, retaining elemental sensory awareness, mind blanking is often described as a state where it seems that consciousness is entirely absent, devoid of both complex thoughts and basic sensory experiences. It is important to note that discerning these states empirically is challenging and that this is an important point to keep in mind if we are to pave the way to disentangle between mind blanking understood as a contentless mental state or as a mental state with a non-linguistically determinable or reportable content.

Van den Driessche et al. (2017) describe mind blanking as a mental state characterized by the absence of reportable content, as opposed to mind wandering, which they present as a content-rich, self-generated, sustained train of thoughts (see also, Kawagoe and Kase, 2020).

It should be noted that this definition of mind wandering as self-generated may be applicable to some examples and not to others, for example, in relation to meditation. Specifically, Delorme and Brandmeyer (2019) discuss meditation as a conscious practice that often involves sustained attention or monitoring of thoughts, and highlight the neurological distinctions between meditative states and ordinary mind wandering. On the other hand, Feruglio et al. (2021) emphasize the varied nature of mind wandering in meditative practices and its potential role in facilitating certain meditation experiences. This comparison underscores the complexity and diversity of mental states; while mind blanking is by some characterized by an absence of reportable content, mind wandering and meditation can both involve self-generated thoughts but differ in intentionality and focus. Moreover, meditation, unlike mind blanking, is typically an engaged process with attentional control and may sometimes incorporate aspects of mind wandering in a more structured and introspective manner.

Interestingly, Van den Driessche et al. (2017) claim that both clinical and subclinical ADHD patients display increased mind blanking, rather than mind wandering. The finding of a metacognitive deficit in ADHD (Antshel & Nastasi, 2008; Knouse, Bagwell, Barkley, & Murphy, 2005), and the well-known association between frontal executive and metacognitive deficits (Shimamura, 1994), concerning self-awareness in particular (Fuster, 1997), suggests that attentional lapses in ADHD may correspond to lower awareness of mental content and thus to episodes of mind blanking. Van den Driessche et al. (2017) aimed to characterize the subjective aspects of inattentive episodes in ADHD. They predicted that people who are hypothesized to have less executive control (e.g., children with ADHD and adults who score high on the DIVA questionnaire) should have decreases in sustained trains of thought. In addition, given the metacognitive deficit, ADHD symptoms might lead to increased reports of mind blanking. What their work highlights is that mind blanking is more likely to occur in individuals who have intrinsic difficulties in sustained attention tasks and in reporting about the content of their mental states (see also, Langland-Hassan et al., 2017). However, this research does not focus on what causes these difficulties in reporting. Instead, this question is explored by Kawagoe, Onoda, and Yamaguchi (2019) in their functional magnetic resonance imaging (fMRI) study. In this research they found that mind blanking's neural correlates may be identified in the deactivation of Broca's area and parts of the default mode network (in particular in the hippocampus) (Christoff et al., 2016) which is, instead, active during mind wandering, in addition to activity in the anterior cingulate cortex in the default mode network. From their neuroimaging data, they conclude that during mind blanking our ability to define the content of our thoughts is impaired because our *inner speech system does not work* at that time. The deactivation of Broca's area, which is associated with language and speech production, appears to constitute the neural representation of mind blanking. When our inner speech system is "silent," we cannot identify the content of our psychological experience. Another possibility, they say, is that we actually *think of nothing* while in a mind blanking state. This interpretation is supported by findings that indicate the deactivation of the hippocampus during mind blanking. Given that the hippocampus is one of the regions showing early activations for spontaneous thought, (as supported by studies on experienced mindfulness practitioners see Ellamil et al., 2016) and that hippocampal damage is linked to a decrease in future and past-oriented thoughts as well as memory (McCormick et al., 2018), participants might actually think of nothing during mind blanking. This hypothesis is reinforced by the fact that the deactivation of Broca's area could be proof of the absence of active thought. Kawagoe and colleagues attribute mind blanking to either metacognitive deficits and/or literally no forms of thinking happening. However, their study did not specifically investigate this aspect, so these two mechanisms might be mixed up in the results.

## 3. Introducing a new methodology

We propose a novel strategy that integrates self-assessment, neuroimaging, and neuromodulation to systematically investigate mind blanking and its neural underpinnings, drawing from approaches employed in the study of other mental states. This combined strategy aims to address the question of whether spontaneous mind blanking (commonly experienced by individuals with ADHD, as per Van den Driessche et al., 2017) and intentional mind blanking (such as experienced by subjects in fMRI studies, according to Kawagoe et al., 2019) share the same neural basis or are supported by distinct neural structures. To evaluate whether participants are experiencing mind blanking as an absence of mental content or as an inability to articulate mental content, we suggest perturbing the brain region thought to be involved in the linguistic expression of mental content - specifically, the left Broca's area - using transcranial magnetic stimulation (TMS) (Epstein et al., 1999; Cattaneo, 2013; Rogič, Deletis, and Fernàndez-Conejero, 2014; Sakreida et al., 2018). This perturbation can be supplemented by employing the task developed by Kawagoe et al. (2019), in which participants are instructed to intentionally attempt to experience mind blanking. In this scenario, the task would be paired with targeted TMS at specific disruptive frequencies, alongside other frequencies or neighbouring areas not involved in the circuit, or placebo stimulation, to serve as control conditions.

In addition to the behavioural changes induced by TMS, cortical alterations that may accompany mind blanking can be detected using electroencephalography (EEG). As mentioned earlier, Andrillon and colleagues (2021) suggested that mind blanking, as manifested in attentional lapses, might be attributed to the occurrence of sleep-like waves, and that regional variations in the occurrence of slow waves appear to predict the nature and behavioural characteristics of such attentional lapses. We intend to incorporate this EEG approach in the context of our experiments.

Furthermore, EEG cannot only capture frequency changes in specific cortical areas during mind blanking, but TMS and EEG can also work in synergy as stimulation and recording techniques respectively. EEG, when integrated with TMS, can be employed to assess the *perturbations* induced in targeted cortical regions. This integration, referred to as TMS-EEG, can be utilized even in the absence of voluntary collaboration of the subject, namely without the need to perform a time-locked task, but only as a probe of frequency or spontaneous behavioural changes (Illmoniemi and Kicić, 2010; Illmoniemi et al., 1999; Illmoniemi et al., 1997). In the protocol we propose, one of the possible probes is to record EEG activity before, during, and after the application of a sequence of repetitive TMS (rTMS) to Broca's area. This protocol is anticipated to facilitate the observation of potential frequency changes associated with episodes of mind blanking (see a similar protocol employed for attention in Herring et al., 2015). Additionally, single-pulse TMS can be employed as a unique tool for probing Broca's area and its connected networks, using EEG as a trial-by-trial readout (by collecting TMS-evoked potentials, or TEPs, both locally and at early latencies) during various mental states experienced by the subject, such as mind blanking, internal dialogue, and mind wandering. This approach has already been extensively used in the study of macro states such as sleep (Massimini et al., 2005; Massimini et al., 2007), anaesthesia (Ferrarelli et al., 2010; Sarasso et al., 2015), and meditation (Bodart et al., 2018; Gosseries et al., 2020).

Regarding the task content, Kawagoe et al. (2019) discovered that participants were capable of intentionally entertaining a state of mind blanking. The experiment was structured in such a way that participants were instructed to make an effort to experience mind blanking. Since these behavioural findings are in line with an earlier report from the same group (see

Kawagoe et al., 2018), the authors posit that the straightforward instructions they employed to induce mind blanking were effective.

In another study, Ward and Wegner (2013) employed measurement techniques based on self-assessment. Instead of instructing participants to attempt to experience mind blanking, they were asked to engage in cognitively demanding tasks, such as reading dense excerpts from Russian literature (specifically, War and Peace and Anna Karenina, provided they were not familiar with these books), and to report instances when they felt they had lost focus on the task. Subsequently, participants had the option to categorize their loss of focus as either mind blanking or mind wandering. The importance of inducing mind blanking in such a spontaneous and natural manner has been recently emphasised (Mortaheb et al., 2022; Boulakis et al., 2023, bioRxiv). The researchers examined both the phenomenological experience and the behavioural consequences of engaging in these reading comprehension tasks and concluded that mind blanking emerged as a unique mental state that occurs spontaneously. They propose that mind blanking possesses a distinct psychological signature in terms of both phenomenology and behavioural outcomes. Additionally, Ward and Wegner investigated the relationship between mind blanking and meta-awareness. Existing research indicates that self-reported and probe-caught accounts of mental states can be used to assess the relationship between a particular mental state and meta-awareness (Schooler et al., 2011). The findings of Ward and Wegner revealed that participants provided self-reported instances of mind blanking, and they further suggest that the relationship between mind blanking and meta-awareness is dynamic over time. In particular, individuals become less aware of mind blanking as time progresses, unless prompted to reflect on their mental states by an external stimulus, such as a request.

We suggest that integrating externally prompted introspection through self-assessment into an experimental strategy could be valuable, especially when complemented with quantitative methods like neuroimaging and neuromodulation.

For these reasons, the disruption (and potential inhibition) of Broca's area using TMS, coupled with EEG responses, could offer two primary advantages: first, it may provide data to help characterize participants' awareness towards the content of their mental states during mind blanking; second, it may facilitate the distinction between spontaneous and intentional mind blanking. Initially, by replicating the experiment of Kawagoe et al. (2019) and instructing participants to attempt mind blanking while applying rTMS to their Broca's area, we could gain insights into the nature of the mind blanking experience, or at the very least, ascertain aspects of it, such as the absence of capacity for inner, conscious speech. Subsequently, we could delve into the possible variations in brain responses to both intentional and spontaneous mind blanking, whether induced by external prompts (as in Kawagoe et al., 2019), during a loss of focus in a demanding cognitive task (as in Ward and Wegner, 2013), or directly induced by rTMS in Broca's area without explicit instructions to mind blank, but which result in comparable behavioural outcomes or compatible EEG patterns.

## 4. Other approaches

Our primary aim in this paper is to suggest potentially promising avenues for future research involving healthy participants, with the possibility to conduct tests and retests under highly controlled conditions, in order to investigate an extremely elusive phenomenon as mind blanking appears to be. However, we would also like to briefly touch upon alternative approaches.

We could explore our research question from the angle of pharmacology. For instance, Van den Driessche et al. (2017) observed that treatment with methylphenidate, a medication commonly used to manage ADHD symptoms, led to a significant decrease in the reports of mind blanking among ADHD patients, effectively aligning them with the levels reported by a healthy control group. Methylphenidate is known to affect neurotransmitter levels (particularly dopamine and norepinephrine). By utilizing neuroimaging techniques, such as functional MRI (fMRI), alongside treatment with methylphenidate, researchers could observe alterations in brain activity patterns and connectivity that correspond to reductions in mind blanking episodes. Additionally, we might consider utilising psychedelic drugs, such as psilocybin or LSD to study the phenomenology of mind blanking. These substances alter consciousness and cognitive processes, allowing researchers to examine their effects on the occurrence and experience of mind blanking. By combining subjective reports with neuroimaging techniques like fMRI and EEG (Carhart-Harris et al., 2016), insights into the neural correlates and cognitive underpinnings of mind blanking during altered states can be explored. Ethical considerations and participant safety should be rigorously accounted for in such studies.

Other pharmacological approaches could involve the use of ketamine and midazolam. For instance, the work of Lehmann et al. (2016) examines the effects of ketamine on the resting state functional connectivity within the default-mode network (DMN) in the context of rumination and distraction. The DMN is known to be associated with internal thoughts, daydreaming, and self-reflection. It could be especially interesting to investigate how the deactivation patterns of the DMN during mind blanking compare to those during ketamine administration. By comparing the resting state functional connectivity under the influence of ketamine with a control condition, it might be possible to determine the neural underpinnings of mind blanking. Furthermore, utilizing different cognitive states like rumination and distraction, as done in Lehmann et al.'s study, could help delineate the boundaries between mind blanking and other mental states.

When considering its application to the study of mind blanking, the sedative and amnesic effects of midazolam might also be of particular interest (see Wang, Sun, and Liang (2020)). Given that mind blanking is a state where individuals are not consciously attending to external stimuli or internal thoughts, the sedative properties of midazolam might induce a similar state. Researchers could compare the brain activity and connectivity patterns during mind blanking with those observed under the influence of midazolam to determine any similarities or differences.

Another approach could involve working with patients who have experienced changes in brain structure or function, such as stroke patients, trauma patients, individuals who have undergone tumour removal surgeries, or patients who have received small coagulations operations to treat their epileptic foci (as in Russo et al., 2021). Specifically, focusing on cases where Broca's area is affected or where patients' experiences resemble mind blanking could be particularly illuminating. By examining the structural and functional brain dispositions through techniques such as MRI, DTI (Diffusion Tensor Imaging) (Jang, 2013), or fNIRS (functional Near-Infrared Spectroscopy) (Strangman, Culver, Thompson, & Boas, 2002) in these individuals, it might be possible to identify any correlations or changes that might occur in the brain during episodes of mind blanking. Moreover, patients' subjective reports can be combined with these data to understand how alterations in Broca's area or its networks affect the phenomenological experience of mind blanking. Lastly, the work of Kucyi et al. (2013) investigating the brain networks involved when the mind wanders away from pain could be relevant. Applying this work to the study of mind blanking, one can explore the interaction between the antinociceptive network, which is involved in pain modulation, and the default mode network, which is active during internal mental processes.

By examining how these networks function during mind blanking, insights can be gained into the brain's ability to control attention and mental content. Furthermore, understanding the neural mechanisms of mind blanking could have implications for pain management strategies by examining its potential role in modulating awareness of pain.

## 5. Conclusion

The current literature does not provide a definitive answer to our question: does mind blanking correspond to a complete lack of mental content or a lack of linguistically or conceptually determinable mental content? We have proposed a research strategy that utilizes TMS and EEG to investigate this phenomenon and identify the neural correlates associated with mind blanking. Additionally, we suggest that differentiating between spontaneous and intentional mind blanking experiences could provide further insight into the underlying neural mechanisms. Based on our analysis of the existing literature, we argue that a multi-method approach, combining self-assessment methods, neuroimaging, and neuromodulation is essential. While we outline a range of approaches, our primary focus is on studying healthy experimental subjects. Specifically, we propose a strategy involving the combined application of TMS and EEG, with an emphasis on monitoring Broca's area activity, in conjunction with self-assessment techniques.

## Conflict of Interest

The authors declare that the research was conducted in the absence of any commercial or financial relationships that could be construed as a potential conflict of interest.

## Acknowledgements

We thank the Editor of the *European Journal of Neuroscience* and the Guest Editors of the special issue "New trends in the empirical study of consciousness: measures and mechanisms" for their support and editorial supervision, and two anonymous reviewers for their constructive comments. We also thank Francesco L. Donati, MD, for his helpful suggestions.

## Author Contributions

The authors confirm being the sole contributors to this work and have approved it for publication.

## List of Abbreviations

ADHD: Attention deficit hyperactivity disorder

DTI: Diffusion tensor Imaging

EEG: Electroencephalography

fMRI: Functional magnetic resonance imaging

fNIRS: Functional near-infrared spectroscopy

TMS: Transcranial magnetic stimulation